\documentclass[9pt,twocolumn]{article}
\usepackage{xcolor}
\usepackage{animate}
\usepackage{graphicx}
\usepackage{subcaption}
\usepackage{siunitx}
\usepackage[super]{nth}
\usepackage{tikz}
\usepackage{amsmath,bm}
\usepackage{amssymb}
\usepackage{physics}
\usepackage[normalem]{ulem}
\usepackage{authblk}
\usepackage[font=footnotesize,labelfont=bf]{caption}
\usepackage[hidelinks,colorlinks=true,linkcolor=blue,citecolor=blue]{hyperref}

\usepackage[symbol]{footmisc}

\def\um{{\micro\meter}}

\title{How a table modulates the risk of airborne transmission between facing individuals}
\date{\vspace{-5ex}}

\author[a]{O\u{g}uzhan Kaplan}
\author[b]{Manouk Abkarian}
\author[a,1]{Simon Mendez\thanks{}}

\affil[a]{IMAG, Univ Montpellier, CNRS, Montpellier, France}
\affil[b]{CBS, Univ Montpellier, CNRS, INSERM, Montpellier, France}

\begin{document}
\twocolumn[
  \begin{@twocolumnfalse}
    \maketitle
    \begin{abstract}
      Airborne transmission has been recognized as an important route of transmission for SARS-CoV-2, the virus responsible for the COVID-19 pandemic. While coughing and sneezing are spectacular sources of production of infected aerosols with far-reaching airflows, the prevalence of asymptomatic transmissions highlighted the importance of social activities. Gathering around a table, a common scenario of human interactions, not only fixes a typical distance between static interlocutors, but influences airborne transmission, by serving both as a flow diverter and a surface for droplet deposition. Here, we use high-fidelity large-eddy simulations to characterize short-range airborne transmission when two people face each other at a typical table. We show that 
compared to the natural distance travelled by free buoyant puffs and jets, the distance between the table and the emission constitutes a new length scale that modifies downward exhaled flows, common during nose breathing, speech, and laughter. When the table is close to the emitter, its main effect is to restrict the forward spread of emitted particles. However, if the distance between individuals is too short, particles reaching the recipient become more concentrated, rising transmission risks. Additionally, simulations of forceful exhalations, like laughter, demonstrate that the table acts as a filter that collects medium-sized particles that would have remained in the free jet otherwise, but can in that case be involved in the fomite transmission route. The table introduces a cut-off size for particles that depends on the inertia of the exhaled material, thereby modifying the size distribution of particles suspended in the air.  
    \end{abstract}
  \end{@twocolumnfalse}
]


\subsection*{Human expiratory flows and airborne transmission}

\footnotetext{*Corresponding author: \href{simon.mendez@umontpellier.fr}{\textcolor{blue}{simon.mendez@umontpellier.fr}}}
More than four years after the onset of the Coronavirus disease 2019 (COVID-19) pandemic, caused by the severe acute respiratory syndrome coronavirus 2 (SARS-CoV-2), the time for emergency is over. 
Nevertheless, the terrible human and economic costs of the pandemic, the possible emergence of similar viruses 
 or new variants of SARS-CoV-2
\cite{Carabelli2023sars}, constitute formidable motivations to elucidate the biological, physical and social mechanisms at play in the transmission of such a disease 
\cite{Wang2021airborne,Galasso2020gender,Zhang2020identifying,Crunfli2022morphological,howard2021evidence}.
 In particular, the question of how microenvironments modulate the transmission risks remains acute~\cite{nielsen2022multiple} to define relevant public policies. 



In airborne transmission of respiratory diseases, susceptible people get infected after inhaling viruses contained in particles suspended in the air, previously emitted by an infected individual
\cite{morawska2020airborne,balachandar2020host,Wang2021airborne,bourouiba2021fluid}. Respiratory droplets are emitted during all expiratory activities, their number and size distribution depending on the activity and on inter-individual factors  
\cite{johnson2011modality,wells1934air,asadi2020effect,Edwards2021exhaled,bagheri2023size}. After their emission, the size of the particles evolves in the air due to the evaporation of their water content.  While leaving the mouth and/or the nose, they are accompanied with a mass of warm, humid air. 
The associated buoyant and turbulent flows modulate the transport of pathogen-laden particles in two principal ways: First, compared to isolated droplets, the settling time of suspended droplets is increased due to their entrapment in the turbulent flow \cite{bourouiba2014violent}. Second, the  evaporation is delayed, the humid and warm expiratory air flows creating a buffer between the droplets and the environment, which is often colder and drier \cite{bourouiba2021fluid,chong2021extended}.

Many studies have been dedicated to coughing and sneezing puff flows from symptomatic individuals \cite{gupta2009flow,bourouiba2014violent,chong2021extended,bourouiba2021fluid}. 
Recently, the importance of presymptomatic, paucisymptomatic, and asymptomatic transmissions of COVID-19 \cite{oran2020prevalence} has highlighted that expiratory modes such as breathing and speech had been overlooked  \cite{stadnytskyi2020airborne,shao2021risk,abkarian2020speech,giri2022colliding,singhal2022virus,Cortellessa2021close,faleiros2022tu}. 
From the flow point of view, breathing and speaking are characterized by successive puffs that may coalesce to form jet-like flows in the far field 
\cite{abkarian2020speech, giri2022colliding,singhal2022virus}.  In addition, flow rates, mouth positions and flow directions vary during speech \cite{abkarian2020speech,Abkarian2020stretching}, possibly generating upward, forward and downward puffs, so that the knowledge gathered for coughs and sneezes is not directly transferable to speech \cite{abkarian2020speech,tang2011qualitative}. 

Superspreading events in indoor settings \cite{Miller2021transmission,azimi2021mechanistic} have shed light on the long-range airborne transmission of SARS-CoV-2, for which exposure times and air renewal rates are critical \cite{bazant2021guideline,morawska2020airborne,bhagat2020effects}.  
On the other hand, in everyday conversations, people may keep a short distance from each other, 
which significantly reduces the time and length scales relevant to the airborne transmission. 
In such `short-range' situations, the fine details of the expiratory air flows and droplet transport are crucial~\cite{Wang2021airborne,nielsen2022multiple}, which has for instance motivated the study of temperature and ambient air currents effects \cite{mendez2023microscopic,Wang2021short-range}. 

However, case-control studies have shown that frequenting places that offer on-site eating and drinking are risk factors for contracting COVID-19 \cite{fisher2020community,galmiche2021exposures}. Indeed, in restaurants, bars and coffee shops for instance, people sit, unmasked, around a table, which fixes a distance of the order of a meter between them. 
The table is a surface on which the largest emitted droplets may deposit, but it also modifies the air flows directed downwards, such as nose breathing, laughs \cite{bhagat2020effects} (see  experimental illustrations in movies S1-S3) and some speech-associated puffs (like `ka' or `ta' \cite{Abkarian2020stretching} or in words ending with a `t', such as `light', for instance). Despite the interest of the table configuration for COVID-19 transmission, the specific studies are rare and do not detail the flow-table interaction~\cite{ding2022infection,faleiros2022tu}. 
In particular, does the presence of a table help or prevent forward propagation? What is the impact of a table for different sizes of respiratory particles? Does it change the concentration in infected particles potentially reaching someone seating opposite an infected person? 

\begin{figure*}[t]
    \centering
	\includegraphics[width=0.85\textwidth]{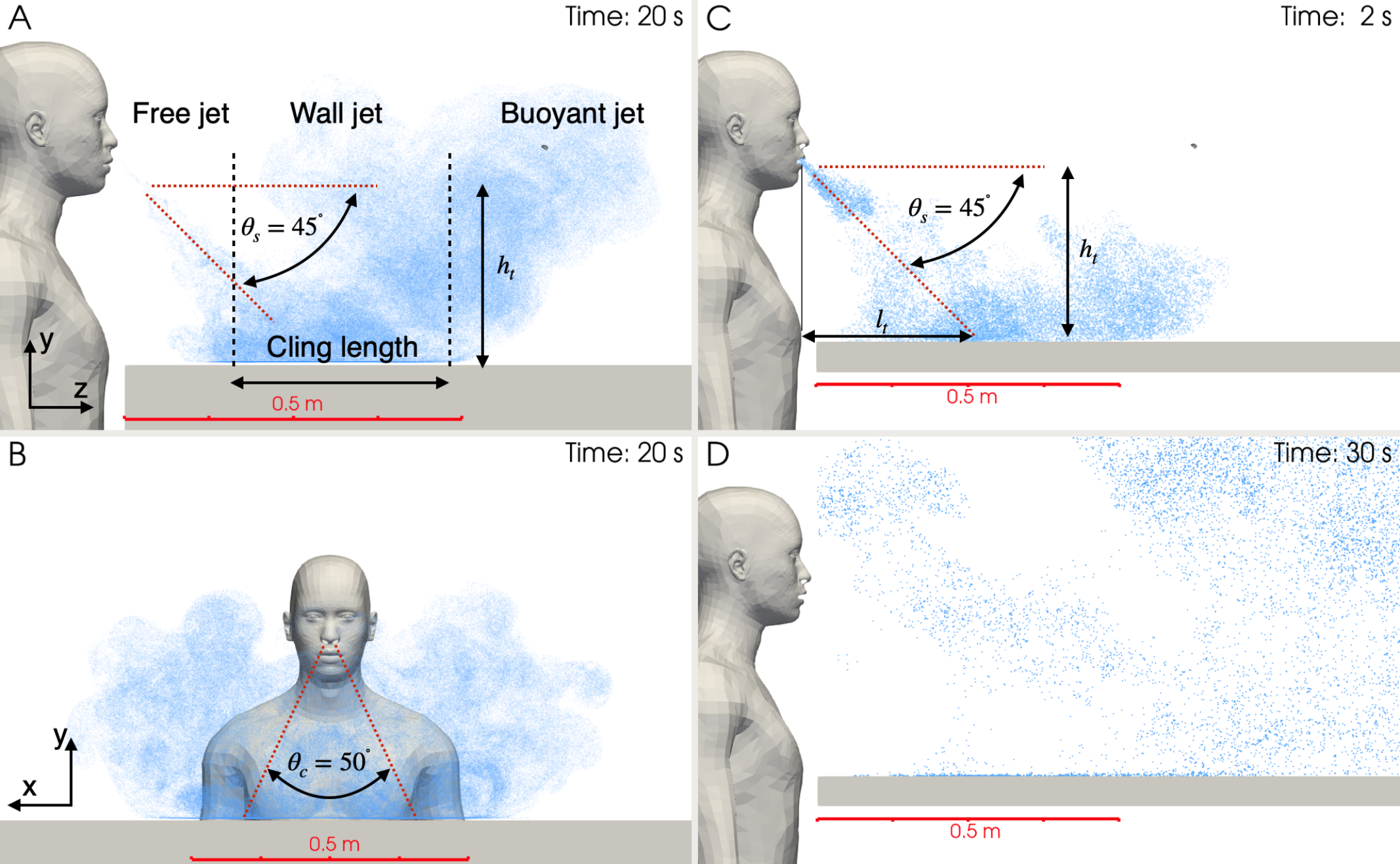}
	\caption{Numerical simulations of expiratory flows considered in the study: a,b) Side and front views of the breathing cloud generated after 5 cycles of periodic nasal breathing ($t=20$ s), for a 15 L/min flow rate and a vertical height between the nostril exits and the table of $h_t=32$ cm. Typical values of orientation of nose breathing jets \cite{gupta2010characterizing} are imposed at the nostrils. 
 c,d) Two snapshots of laughing flow in front of a table at $t=2$ s (end of the injection signal) and $t=30$ s, respectively. A $45^\circ$ downward orientation is imposed.  Particles are single-colored. Note the large structures at the head of the jets, similar to observations of Ref. \cite{sharp1977buoyant} and \cite{burridge2017free}. The flow direction, impact on the table, clinging, lateral spreading in nose breathing are qualitatively illustrated in experimental movies S1-S2 (nose breathing) and S3 (laughing).
}
	\label{fig:config}
\end{figure*}

In this paper, we use numerical simulations of expiratory flows to explore those open questions. Focus is made on two configurations chosen to mimic archetypes of expiratory flows: nose breathing, in which periodic exhalations coalesce to form a jet interacting with the table, and laughter, with a sudden series of air expulsions forming a puff impacting on the table, which is also a model for coughs or strong speech flows, such as plosion utterances.
The scenarios are illustrated in Fig. \ref{fig:config}.

\subsection*{Breathing Flow}
Nose breathing is first considered. During such a maneuver, emitted particles are less than a few microns in diameter and originate from the lower respiratory tract \cite{bagheri2023size}. 
Their small size makes them readily transported by the exhaled flow and suspended for long times in the surrounding environment \cite{morawska2020airborne,Netz2020physics}. 
The impact of evaporation on their trajectory is negligible. 
As a consequence, the trajectory of the exhaled particles is determined by the flow itself, which justifies focusing on the jet dynamics. 
Nose breathing flows are mainly characterised by three features \cite{gupta2010characterizing}: 1) the flow rate is roughly periodic due to the successive exhalations and inhalations; 2) the flows are directed downward, with an angle in both the lateral and the vertical directions; 3) the exhaled air is generally warmer than the ambient environment. First, as observed for mouth breathing \cite{abkarian2020speech}, we can expect that the periodic exhalation and inhalation only matters close to the nose, the puffs of the successive breaths coalescing to form two starting jets (one at each nostril). The initial vertical momentum of the warm jets is opposed to the body forces caused by the density difference between the exhaled air and surrounding environment, which makes the jets negatively buoyant 
\cite{turner1966jets,fischer1979mixing}. This is well visualised in Fig.~\ref{fig:config}A,B, where the exhaled jets descend, then lift, propagating over a horizontal distance of approximately one meter. The inertial and thermal effects can be quantified by the Reynolds number \( Re=U_0 D_0/\nu \) and the densimetric Froude number \(Fr=U_0/\sqrt{g' D_0} \), with
$U_0$ the time-averaged bulk velocity, $D_0$ the hydraulic diameter of a nostril, {$D_0=\sqrt{4S/\pi}$, with $S$ its surface area},
$\nu$ the kinematic viscosity of air and $g'$ the reduced gravity due to the density difference between the jet and the ambient fluid.

Nose breathing flow is investigated using $D_0=1.23$ cm, an exhaled volume rate of 15 L/min \cite{gupta2010characterizing} and temperatures of $T_{\text{exh}} = \SI{32}{\degreeCelsius}$ for the exhaled air and $T_{\text{amb}} = \SI{25}{\degreeCelsius}$ for the ambient air (see the Materials and Methods). This yields $Re=1450$ and $Fr=25$.
A cycle of 4 s with  1 L exhaled per breath  is repeated 20 times to compute 80 s of physical time (Fig. S1 shows the signal used). 
The jet is seeded at constant concentration with massless tracer particles, whose dynamics is essentially the same as the small droplets exhaled during nose breathing, allowing to characterize the cloud generated by the nose breathing along time.
Using the residence time of tracers $t_r$, which denotes the time since their injection, an average trajectory $\bm{\vec{X}}(t_r)$ can be calculated as
\vspace{-10pt}

\begin{equation}
    \bm{\vec{X}}(t_r)=\frac{1}{2N}\sum_{j=1}^{2} \sum_{i}^{N} \vec{x}_{i,j}(t_r),
    \label{eq:trajectories}
\end{equation}
\vspace{-10pt}

\noindent  $\vec{x}_{i,j}(t_r)=(x,y,z,t_r)$ is the coordinates vector of the $i$th tracer emitted from nostril $j$, $N$ the number of tracers injected at each nostril. The symmetry of the geometry with respect to $x$ yields an average trajectory in the ($z$,$y$) plane (see Fig. \ref{fig:config}). 

\begin{figure}[tbhp]
	\centering
	\includegraphics[width=0.75\linewidth]{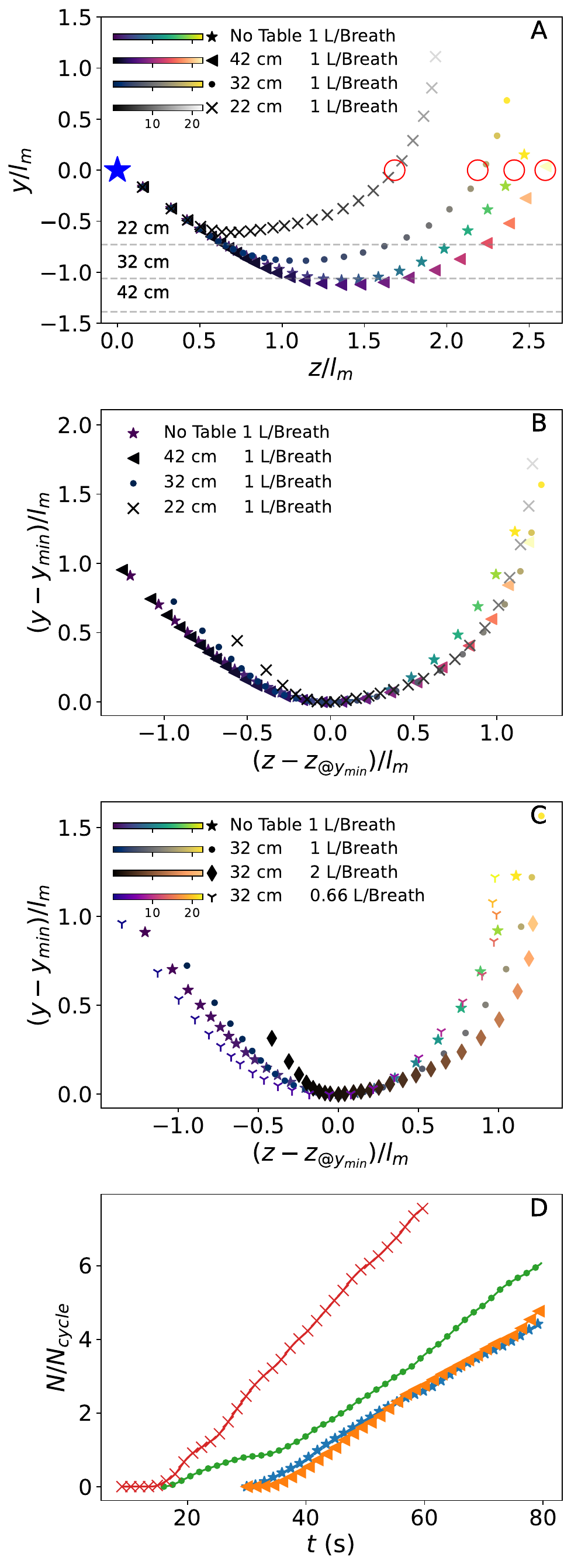}
	\caption{Average trajectories of breathing jets, color-coded by residence times of the massless tracers normalised by $l_m=D_0 Fr$. (a) Cases with and without table, for a flow rate of 1 L/breath ($Fr=25)$; (b) Same trajectories with the origin shifted at the location of minimum height at the jets ($z_{@y_{min}}$;$y_{min}$); (c) Same plot as (b) assessing the effect of the flow rate (or Froude number). (d) Number of tracers entering a spherical probe of radius 5 cm placed where the averaged trajectory returns to the emitter elevation, indicated by red spheres in (a). $N_{cycle}$ denotes the number of particles injected per breath, which remains the same for the presented cases. In (a), horizontal dashed gray lines indicate vertical distance of the tables from the mouth of the emitter, denoted by a blue $\star$. }
	\label{fig:trajectory_breathing}
\end{figure}

The trajectory of the unbounded jet (no-table), is shown in Fig.~\ref{fig:trajectory_breathing}A (see also Fig. S2). Here we introduce the classical length scale $l_m=D_0 Fr$ to nondimensionalize the coordinates, which is further discussed below. 
The inclined, unbounded jet is initially momentum-driven and its trajectory first follows a straight line. As the jet gradually expands in the cross-stream direction, it  decelerates due to both the entrainment of the ambient fluid and the opposing buoyancy force. Eventually, the buoyancy overcomes the vertical momentum flux, causing the jet to reach a terminal depth at $y\approx l_m$ before going upwards, along a skewed trajectory that is best characterised by quartic polynomial (\cite{papakonstantis2020simplified}, see also Fig. S2). The length scale $l_m$ indeed characterizes this transition from jet-like to plume-like motion \cite{roberts1997mixing}. It reads $l_m = D_0 Fr$ \cite{fischer1979mixing} (see the Materials and Methods). For the jet considered here, the mean value for $l_m$ is found to be approximately 30~cm.

Clinging results from a `Coanda-like effect' in the lower part of the jet, where the limited entrainment of the ambient air because of the wall {creates a vertical pressure gradient,  pulling the jet toward the surface.}
Sharp and Vyas explored the conditions that lead to the clinging of a jet (refer to Fig 5. in \cite{sharp1977buoyant}). In the present case, clinging is estimated to occur for tables placed less than 48 cm (1.6 $l_m$) lower than the nose. 
Simulations of the interaction between breathing jets and a horizontal table are presented with three nose-wall separation heights: $h_{t}=22$, $32$, and $42$ cm {(which correspond to 0.73 $l_m$, 1.06 $l_m$ and 1.39 $l_m$, respectively)}. Clinging is thus expected for the three cases \cite{sharp1977buoyant}. 

In Fig.~\ref{fig:trajectory_breathing}A, the average trajectories of the `table-jets' are compared to that of the free jet. 
Initially, the table-jets develop as the unbounded jet. Then, the  interaction with the table results in the formation of buoyant wall jets (see Fig.~\ref{fig:config}A and Movie S4). When the table is  higher than the terminal height of the free jet trajectory (cases 22 cm and 32 cm), the jet development is strongly constrained, yielding a shorter penetration in the vertical and the horizontal directions.
For lower table heights, the jet-wall interaction is less pronounced. 
Nevertheless, the convergence to the free-jet trajectory is not monotonic, as shown by the 42-cm case: the jet trajectory barely differs from the free jet in the descending region. However, due to clinging, the wall prevents the jet from lifting normally due to buoyancy forces \cite{sharp1977buoyant}, so that the jet is closer to the wall and forward penetration near the table is enhanced. 

In order to compare the trajectories of the wall jets, they are plotted with their origin shifted to the location of the minimum vertical position  ($z_{@y_{min}}$;$y_{min}$) in Fig.~\ref{fig:trajectory_breathing}B. It shows that the three wall jets trajectories actually superimpose in the ascending portion, 
even though the jets have interacted with the table at different distances from the source, and thus different velocity, diameter, Froude number: in particular, friction forces are different, as higher velocities are obtained closer to the wall when $h_t$ is small: the decrease in the forward motion is thus not related to viscous friction.  
The interpretation is that in the  jet initial development, its diameter $D(s)$ increases like $s$, the distance from the emission and the local Froude number $Fr(s)$ decreases like $1/s$, so that the quantity $D(s) Fr(s)\approx D_0 Fr = l_m$ is conserved at the jet-wall impact. Thus, even if the wall jets have different ratios of momentum and buoyancy forces, the characteristic length needed for the buoyancy forces to overcome the momentum forces is conserved, which yields similar ascending trajectories. 


To explore the effect of exhaled volume on the 
propagation of breathing clouds, we use simulations for three different exhaled volumes (0.66, 1.0 and 2.0 L per breath, corresponding to approximately $Fr=17$, $25$ and $50$, respectively) at constant $h_{t}=32$ cm (Fig.~\ref{fig:trajectory_breathing}C). The jet with the lowest breathing volume does not cling and behaves similarly to a free jet. However, for the clinging jets, the trajectory depends on the Froude number. For the largest breathed volume (2 L/cycle), a notable increase in the near-wall horizontal extent of the jet is observed. This is consistent with the non-linear behavior of cling length with $Fr$ predicted by Sharp and Vyas \cite{sharp1977buoyant}). In cases with high Froude number, the trajectory near the wall becomes flatter due to the mitigation of buoyancy effects.

In turbulent jets, the concentration of the  discharged material decreases as the inverse of the distance from the emission $1/s$. 
Let us consider the height of the emitter's nose ($y=0$) as the height at which infectious particles may be inhaled by an interlocutor. Fig.~\ref{fig:trajectory_breathing}A  shows that
 `table-jets' at $h_{t}=22$ and $32$ cm return to $y=0$ by travelling a shorter distance than the free jet. As a consequence, they are expected to carry a larger density of infectious particles to a susceptible person reached by the jets. 
To quantify such an effect, we place fictive spherical probes of 5 cm radius at the positions of the jet-centerline where they reach $y=0$, as shown by red markers in Fig.~\ref{fig:trajectory_breathing}A. For each jet, we report the number of particles that have entered the spherical probe as a function of time (Fig. \ref{fig:trajectory_breathing}D), which provides a cumulative number of particles that may reach a susceptible interlocutor along time. 
First, differences between the free jet and the `42 cm' table jet are small.
On the contrary, transmission risks are increased in the `22 cm' and `32 cm' table-jets by two effects: first, the time at which the first particles reach the interlocutor's face is smaller than for the free-jet; then, the slope of the curves, which informs about the particle flux, indicate that the local particle concentration opposite to the emitter is more intense in the table jets when $h_{t}/l_m \lesssim 1.0$ (22 and 32 cm cases). 

\begin{figure*}[t]
    \centering
	\includegraphics[width=\textwidth]{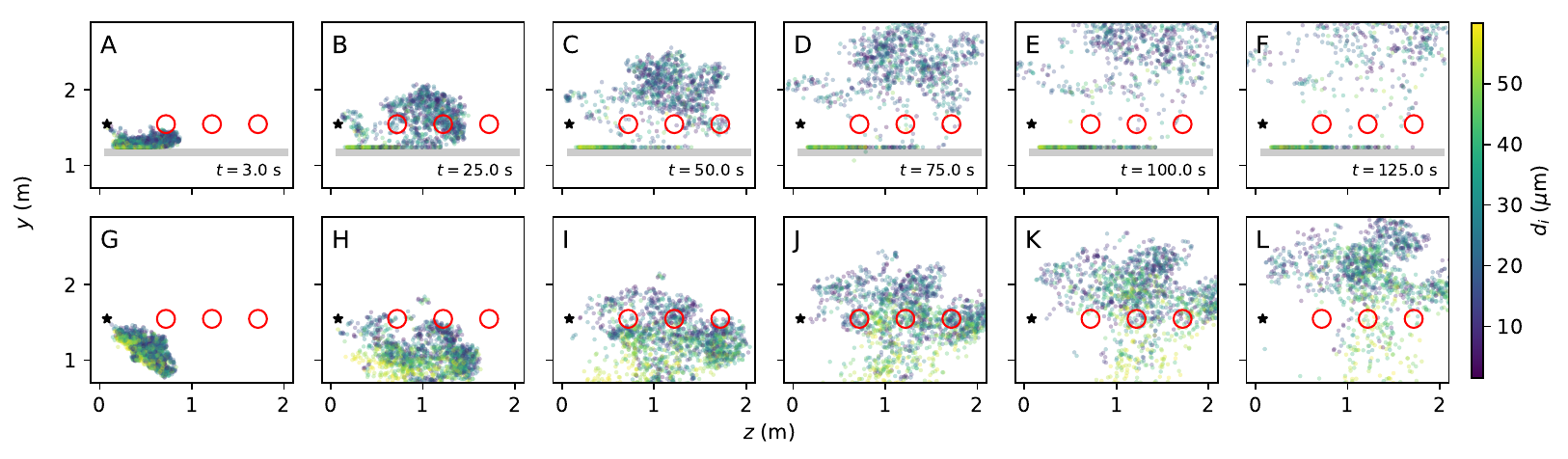}
 	\vspace{-20pt}

 \caption{Snapshots of particle clouds emitted during the laughing flow at 6 different time instants: configurations with the table (top row, A-F) and without the table (bottom, G-L). Each column shows the same time instant. Droplets are colored by their initial diameter. The \textcolor{black}{$\star$} indicates the mouth of the emitter. Three spherical probes used for evaluating the particle fluxes at the emitter/receiver face height are displayed by \textcolor{red}{$\circ$} (see Fig \ref{fig:fig4}A). They are located at 70 cm, 120 cm and 170 cm from the mouth exit and mimic different separation distances between two people seating at a table. The computational domain is larger than the viewing window.}
	\label{fig:particle_cloud_laugh}
\end{figure*}

\subsection*{Laughing flow}
While breathing flow is characterised by periodic exhalations at moderate Reynolds numbers, some of the human exhalation types are shorter and faster: shout, cough, laughter, sometimes speech, not to mention sneeze. We consider here a signal mimicking laughter, composed of 2 series several short and high-speed exhalations directed downwards as in laughter (see Fig. S1). Laughter can indeed generate high-speed downward flows \cite{bhagat2020effects}, as illustrated by an experiment shown in movie S3 (and Fig. S5). 

An inflow signal of 1.5 s per cycle is considered, with a flow rate profile that captures the intermittent feature of laughing flows with high-frequency bursts generating coalescent puffs: it consists of an exhalation of 1 s with 5 peaks at 2 to 3 L/s and an inhalation of 0.5 s. Two consecutive cycles are computed. The total volume exhaled is 1 L/cycle (see Table S1).
The inflow is issued from a mouth in elliptic-like shape. 

The maximum Reynolds number during the emission is approximately 11750.
The flow is directed forward and downward, at an angle of $\SI{45}{\degree}$ with respect to the streamwise axis $z$, and has a temperature of $T_{exh}= \SI{34}{\degreeCelsius}$ with relative humidity (RH) of 90$\%$. The flow is released into a still environment. Particles with initial diameters ranging from \SI{1}{\micro\metre} to \SI{60}{\micro\metre} with a uniform distribution are injected at the inflow with an initial velocity equal to that of the gas. These particles are used as a proxy for expiratory droplets and  evaporate down to a minimum diameter. 
The evolution of the particle cloud is studied in the presence of a horizontal table located 40 cm lower than the person's mouth (Fig.~\ref{fig:config}C,D), or without that table.
In both cases, the emission lasts less than 3~s, but the simulation is continued without any input flow up to $t=$ 125~s to study the particle cloud and associated flow dynamics.

Fig.~\ref{fig:particle_cloud_laugh} shows the evolution in time of the particle cloud in both configurations. In the `no-table' case (bottom row), the  cloud is first dominated by inertia. 
When buoyancy  overwhelms the vertical momentum, the cloud is deflected 
upwards, as shown by the dynamics of the smallest particles. The largest droplets  leave the cloud early in the descent stage and are mostly settled out of the cloud during its ascent. 
As in the breathing flow cases, when a table is placed above the location of the terminal descent of the free jet, the exhaled flow impacts the table, which cancels its vertical momentum and limits its forward motion (top row). 
The maximum streamwise distance reached in the `no-table' configuration is more than 2.5 m, but less than 2 m in the case with the table.
A number of particles deposit on the table, which  will be analyzed further. 

In the table case, the ascending phase starts almost immediately after the puff injection (see Fig.~\ref{fig:particle_cloud_laugh}A, $t=3.0$ s). This early deflection makes the puff rapidly reach potential interlocutors  in front of the emitter. 
As the distance travelled from the source $s$ is proportional to $t^{1/4}$ and that the concentration field evolves as $1/s$ \cite{sangras2002self} in free puffs, we estimate that concentration varies with time as $c\approx t^{-1/4}$. Therefore, we predict that the `table-puff' may carry a higher droplet concentration than free-puff would when reaching a receiver. 

\begin{figure}
    \centering

	\begin{subfigure}{\linewidth}
    \hspace{-5pt}
	\includegraphics[width=\linewidth]{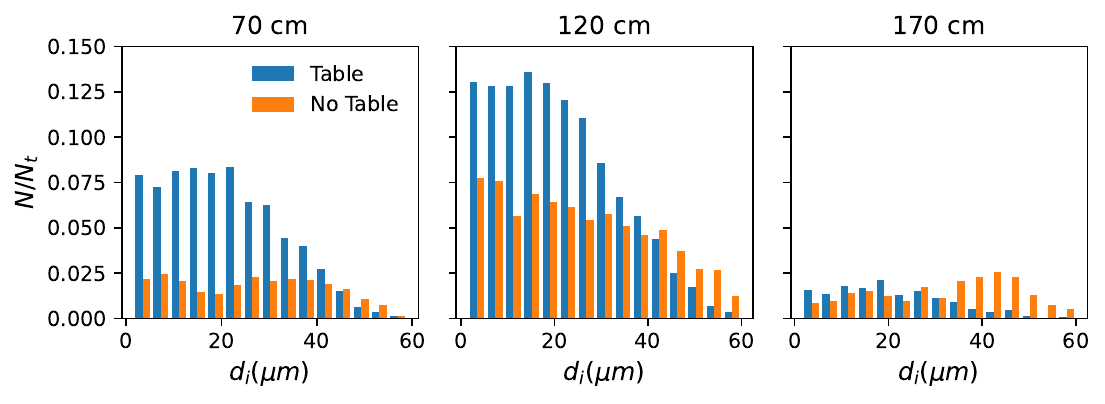}
		\label{fig:particle_dist_laugh}
	\end{subfigure}
	\begin{subfigure}{\linewidth}
	\includegraphics[width=\linewidth]{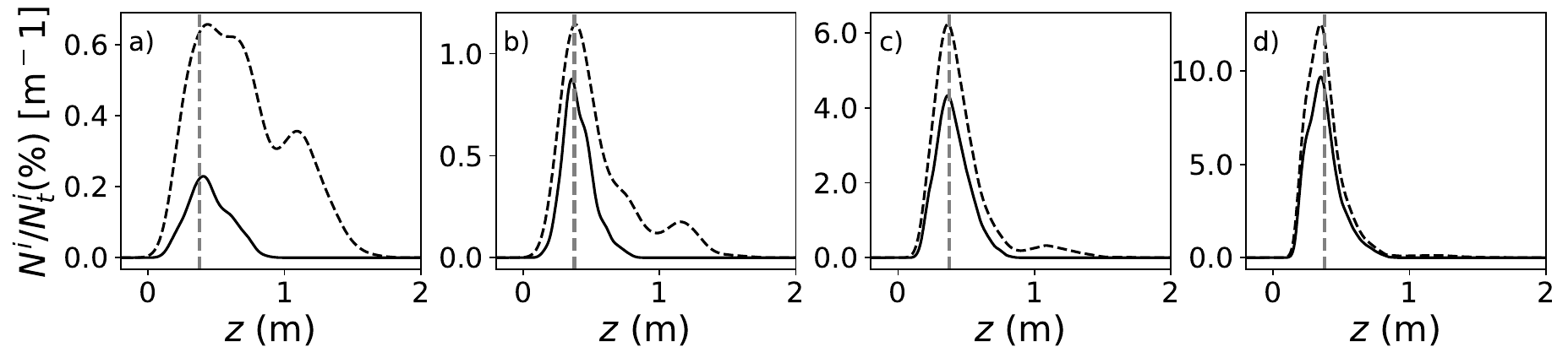}
	\label{fig:particle_table_laugh}
	\end{subfigure}

 \begin{tikzpicture}[overlay]
        \node[above left] at (0.235\textwidth,0.29\textwidth) {A};
        \node[above left] at (0.235\textwidth,0.125\textwidth) {B};
    \end{tikzpicture}

	\vspace{-25pt}
	\caption{Laughing flow: Airborne (A) and deposited (B) particles. (A) Fraction of particles passing through spherical probes centered 70, 120, 170 cm away from the emitter (red circles in Fig.~\ref{fig:particle_cloud_laugh}) during the  $125$ s of simulation, as a function of their initial diameter. (B) Case with table: density of particles deposited on the table as a function of the streamwise coordinate $z$. Approximately \SI{33000} particles are injected, with a uniform distribution between \SI{1} and \SI{60}{\micro\meter}. Particles are grouped into 4 bins based on their initial diameter, i.e a) \SI{1}{}-\SI{15}{\micro\metre}, b) \SI{15}{}-\SI{30}{\micro\metre},  c) \SI{30}{}-\SI{45}{\micro\metre}, d) \SI{45}{\micro\metre}-\SI{60}{\micro\metre}. Solid and dashed lines denote 5 s and 125 s of simulated time, respectively. The vertical dashed line is the location of the geometric impinging point of the puffs.}
  \label{fig:fig4}
\end{figure}
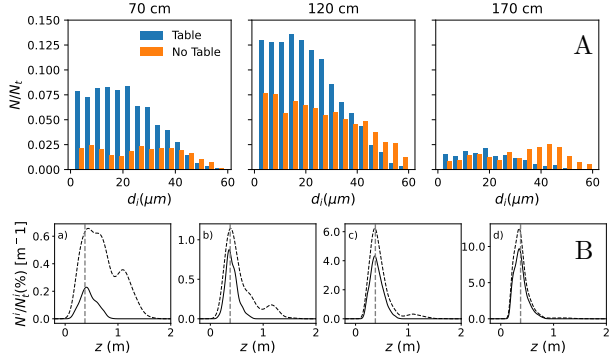


To quantify the impact of the table on the particles that may be inhaled by a person located in front of the  emitter at different distances, three spherical probes with a radius of \SI{5}{\centi\metre} are placed  opposite the emitter at 70, 120 and 170 cm. Fig.\ref{fig:fig4}A displays the histogram of diameters for the particles  having entered those spheres during the simulation. 
At short distances, the puff-table interaction significantly increases the number of particles with an initial diameter less than \SI{40}{\micro\metre}, compared to the case without table. However,  few particles reach the probe at $z=170$ cm. 
On the other hand, for the three locations considered, the number of droplets with an initial diameter larger than \SI{40}{\micro\metre} is larger without a table: in that case, flow motions counteract sedimentation and transport some of the largest droplets simulated to a height where they may be inhaled. 
The deposition of particles on the table is now examined to understand this difference. 


In Fig.~\ref{fig:fig4}B, we present the density in the $z$ direction of particles deposited over the table, integrated over the spanwise direction $x$. (The integral over $x$ and $z$ is provided in Fig. S3.) Particles are grouped according to their initial diameter and results are plotted at 5 s, soon after the injection, and at 125 s. 
First, the different scales on the y-axes indicate that the proportion of deposited particles increase with their size (see also Fig. S3). Moreover, particles within the \SI{30}-\SI{60}{\um} range rapidly deposit, as evidenced by the small differences between the 5 s and the 125 s curves: large particles actually impact on the table due to their initial inertia. 
The maximum deposition of these droplets occur near the geometric stagnation point, which is consistent with the deposition profile in impinging jets with large nozzle-to-surface distances \cite{burwash2006deposition}. 
In contrast, the deposition of smaller particles is more progressive in time and spans over a larger region: it is associated with their sedimentation from the cloud. 
Fig.~\ref{fig:fig4}B gathers results from the two cycles simulated. 
Differences between the two cycles are actually obtained for the smallest particles: the particles emitted during the first cycle 
are pushed further downstream and contribute more significantly to the deposition for $z \gtrapprox 0.8$~m (see Fig. S4). Overall, the table acts as a filter that collects the largest particle at impact, which modifies the distribution of airborne particles and explains the absence of large airborne particles observed in Fig.~\ref{fig:fig4}A in the table case.

\subsection*{Conclusions}


We have used 
large-eddy simulations to explore human expiratory flows and the associated particle transport around a table, to understand short-range airborne transmission from asymptomatic individuals 
in everyday social interactions, for example, in meetings and restaurants. 

Different types of exhalations generate flows directed downwards which, in a colder medium, first have a straight trajectory before being deflected by buoyancy forces. If a table is located above the terminal depth reached by the puffs or jets, it obstructs the vertical penetration of flows and causes them to deviate earlier. While this shortens the forward propagation of the exhaled material, the flows also reach a susceptible person facing the emitter sooner since their injection, thus at higher concentration,  potentially increasing pathogen exposure and transmission risk. We have shown that the presence of a table introduces a new length scale $h_t$, the distance between the nose/mouth of the emitter and the table, which modifies the free flow dynamics as soon as it is shorter than the 
terminal penetration of the flow without table. 

Finally, not only the table modifies the trajectory of the exhaled material, but also the content of the exhaled clouds. Indeed the impact of the flow on the table makes it an inertial impactor \cite{marple1974characteristics} that collects the largest particle. The table thus
introduces a cut-off diameter for the particles able to remain suspended and participate to airborne transmission. 
As the response of inertial impactors is known to scale as the square root of the particle Stokes number \cite{marple1974characteristics}, one expects that the cut-off particle diameter for impaction on the table evolves like the inverse of the square root of the inflow velocity: the faster the exhalation, the smaller the particles remaining in the air after impact. Interestingly, this also allows to provide new insights in the particles that would participate to the fomite transmission route after deposition on the table surface. 

Like most short-range transmission studies \cite{chong2021extended,abkarian2020speech,Wang2021short-range}, we investigated how an infected individual seeds the immediate surroundings with infectious particles, here in the presence of a table, and use airborne concentration as a proxy to transmission risk. The thermal boundary layer (TBL) due to the body temperature of the emitter has been neglected, the exhaled air velocities at the mouth/nose exit being much larger than those in the TBL, in particular in the presence of a table  \cite{Licina2014experimental}. 
On the contrary, it would be necessary to account for the TBL of a susceptible interlocutor \cite{giri2022colliding,singhal2022virus}  if they were explicitly included in the simulation. Indeed, including a second person in the simulation would be especially interesting in the cases where the emitted flow is sufficiently energetic to travel along the whole table and impact the second body. In that case, we may expect the transport of the airborne particles from the table to the person's face to be influenced by their TBL.


By demonstrating the richness of the phenomena at play in a simple configuration, our study opens the way to numerous possibilities in short-range airborne transmission research, for instance by coupling geometrical, flow and thermal effects. In addition, it  highlights the interest of investigating more fundamental situations, as the dynamics of impacting turbulent puffs, or unsteady effects in inertial impactors. 
We are only at the beginning of the investigation into the effects of geometry settings on the transmission of respiratory diseases. 

\subsection*{Materials and Methods}
{Large-eddy simulations of the variable-density Navier-Stokes equations in the low-Mach regime 
were performed using the in-house flow solver YALES2BIO, already used in the airborne transmission context \cite{abkarian2020speech,mendez2023microscopic}. Droplets are accounted for by a one-way spherical point-particle formalism accounting for drag and gravity forces. Their diameter evolve due to evaporation. They are assumed to be made of water, but include a non-volatile fraction that limits their evaporation: the final diameter is 1/3rd of the initial diameter, which corresponds to a non-volatile fraction of 3.7\% \cite{Netz2020physics}.

Characteristics of human nasal respiration 
vary with various factors, such as anatomy, age, posture and physical activity, \textit{etc.} \cite{gupta2009flow}. 
Here, we do not aim to examine all possible flow conditions: 
we rather use values that are typical of breathing and explore the effect of some parameters, focusing on the interaction with a horizontal table. We define a 4 s cycle, with an exhalation time of 2.4~s (see Fig. S1).
We set that jets are inclined by $\SI{45}{\degree}$ downwards with respect to the streamwise axis $z$ and and to \SI{25}{\degree} to the symmetry plane (see Fig. \ref{fig:config}). Breathing volume per breath is varied from 0.66 L to 2 L per breath, which is a heavy breathing. The hydraulic diameter of each nostril is ${D_0=\smash{\sqrt{4S/\pi}}=1.23}$ cm, with its surface area $S$=\SI{1.18}{\square\centi\metre}. 20 cycles are computed, for a total physical time of 80 s simulated. For Eq. \ref{eq:trajectories}, we consider particles with injection times $t_i$ ranging from 20 to 80 seconds to obtain well-converged trajectories from $t_r=0$ to \SI{25}{s}, discarding the initial transient phase. 

For the laugh flow, a 2-cycle signal previously used for speech \cite{abkarian2020speech} is compressed in time so that {the exhalation in each cycle} lasts 1 s and the exhaled volume per cycle is 1 L  (see Fig. S1). The flow exits from an elliptic mouth surface of hydraulic diameter \SI{2.46}{\centi\metre}. After the 2 cycles of injection, a large puff is formed. The simulation is continued until \SI{125}{s} without any flow rate at the mouth. 

For the laugh flow, we have used a temperature of $34^\circ$C, classical in human exhalation studies. For the nose breathing, a temperature of $32^\circ$C has been used, to render the heat exchange role of the nose. However, the absolute values are only indicative, the important figures being obviously the non-dimensional numbers that quantify the relative effects of the different physics involved.  Relative humidity has been set to $90\%$ in the exhaled air and $50\%$ in the ambient air, whose temperature is fixed to $25^\circ$C. A more detailed characterization of the flows are provided in the SI.

The length scale $l_m$, representing the transition from jet-like to plume-like motion, is derived through dimensional analysis as $l_m$ = $M_0^{3/4}/B_0^{1/2}=D_0 Fr$. $M_0$, $B_0$ are the specific fluxes of momentum and buoyancy \cite{fischer1979mixing}, defined as \(Q_{0} = \pi D_0^2 U_0/4, M_0 = Q_0 U_0, B_0 = g'Q_0, g' = g{\abs{\rho_{amb}-\rho_{exh}}}/{\rho_{amb}}, Fr={U_0}/{\sqrt{g'D_0}}\), with $Fr$, $U_0$,  $g'$ and $D_0$ the densimetric Froude number, the jet velocity, the reduced gravity and the nostril diameter, respectively. 

}
\subsection*{Acknowledgments}
{The setup of the CFD simulations was designed with P. B\'enard, G. Lartigue, V. Moureau (CORIA, France), G. Balarac, P. B\'egou (LEGI, France), Y. Dubief (Univ. Vermont, USA) and R. Mercier (Safran Tech, France). CFD simulations were performed using HPC resources from TGCC-IRENE (Grants No. A0100312498 to A0140312498). This work was funded by the Agence Nationale de la Recherche, project TransporTable (ANR-21-CO15-0002) as well as by the LabEx NUMEV (ANR-10-LABX-0020) under the SATIS project, within the I-Site MUSE (ANR-16-IDEX-0006).}

\bibliographystyle{unsrt}
\bibliography{covid_biblio}

\end{document}



\maketitle



\section*{Simulation methodology}


Variable-density simulations are considered in this paper \cite{moureau2011design}. This allows to account for buoyancy effects and evaporation of the emitted droplets.
In addition to momentum and mass conservation equations for the air, droplets drag, mass and energy transfers are also considered \cite{boulet2018modeling}. 
The aim of this section is to provide more details about the simulation methodology described in the Materials and Methods section of the article.

\subsection*{Physical models}

\subsubsection*{Gas phase}

We focus on the flow of hot and humid exhaled air discharged in an ambient air at rest. Air is represented as a mixture of five species: nitrogen, oxygen, carbon dioxide, argon and water vapor. Ambient and exhaled air are modeled as the same fluid with a different composition in the different gases, the exhaled air containing more carbon dioxide and water vapor, in particular. 

The typical flow velocities considered in this study (a few meters per second) yield low-Mach number flows, in which compressibility effects can be neglected. The flow is therefore the solution of the Navier-Stokes equations in their variable density form. In Large-Eddy Simulations, the unknowns are the spatially filtered physical unknowns representing the large scales of the turbulent flow.
In the following, $\,\bar{}\,$ and $\,\widetilde{}\,$ symbols will denote the unweighted and Favre filtering operators, respectively. The filtered  equations for the conservation of mass, momentum, energy and species are, respectively:
\begin{equation}
\label{eq:mass_conservation}
\frac{\partial \overline{\rho}}{\partial t} + \frac{\partial \overline{\rho} \widetilde{u}_i}{\partial x_i} = \sum_{k=1}^{n_{sp}}\Theta_{M,k},
\end{equation}

\begin{equation}
\label{eq:momentum_conservation}
\frac{\partial \overline{\rho} \widetilde{u}_j}{\partial t} + \frac{\partial \overline{\rho} \widetilde{u}_i \widetilde{u}_j}{\partial x_i} = -\frac{\partial \overline{P}}{\partial x_j} + \frac{\partial (\tau^{lam}_{ij} + \tau^{sgs}_{ij})}{\partial x_i} + \Theta_{D,j},
\end{equation}

\begin{equation}
\label{eq:energy_conservation}
\frac{\partial \overline{\rho} \widetilde{h}_s}{\partial t} + \frac{\partial \overline{\rho} \tilde{u}_i \widetilde{h}_s}{\partial x_i} =  \frac{D\overline{P_0}}{Dt} + \frac{\partial}{\partial x_i}\left(\lambda \frac{\partial \widetilde{T}}{\partial x_i} - \frac{\mu_t}{Pr_t}\frac{\partial \widetilde{h}_s}{\partial x_i}\right) + \frac{\partial}{\partial x_i} \left(-\overline{\rho} \sum_{k=1}^{n}D_k \frac{\partial \widetilde{Y}_k}{\partial x_i} \widetilde{h}_{s,k}\right) + \Theta_{h_s},
\end{equation}

\begin{equation}
\label{eq:species_conservation}
\frac{\partial \overline{\rho} \widetilde{Y}_k}{\partial t} + \frac{\partial \overline{\rho} \widetilde{u}_i \widetilde{Y}_k}{\partial x_i} = \frac{\partial}{\partial x_i}\left(\frac{\overline{\rho} \mu_t}{Sc_t}\frac{\partial \widetilde{Y}_k}{\partial x_i}\right) + \frac{\partial}{\partial x_i}\left(\overline{\rho}D_k \frac{\partial \widetilde{Y}_k}{\partial x_i}\right) +  \Theta_{M,k},
\end{equation}



\noindent where the density, the velocity, the dynamic pressure, the temperature, the sensible enthalpy and the mass fraction of the species $k$ are denoted by $\rho$, $u$, $P$, $T$, $h_s$ and $Y_k$. The dynamic viscosity, the molecular diffusion coefficient of species $k$ and the thermal conductivity are denoted by $\mu$, $D_k$ and $\lambda$. The conductive heat flux is modeled according to Fourier's law, as proportional to the filtered temperature $\widetilde{T}$ gradient, with the thermal conductivity $\lambda$ as proportionality coefficient.
The energy conservation is written here with the sensible enthalpy $h_s = \int_{T_0}^T C_p \, dT$, with $C_p$ the heat capacity at constant pressure. To close the problem, the ideal gas law for filtered quantities is supposed to hold: $\overline{P_0} = \overline{\rho} \widetilde{r} \widetilde{T}$, where $P_0$ is the thermodynamic pressure and $r$ is the mixture gas constant.
The $\Theta$ terms are source terms mimicking the effects of the dispersed phase on the fluid. In the present study, a one-way approach is used, so that the $\Theta$ terms are all neglected. The effect of the gas phase on the particles is accounted for, but there is no feedback. This is motivated by the low concentration of emitted droplets during the maneuvers simulated \cite{bagheri2023size}.

In Eq.~[\ref{eq:momentum_conservation}], the filtered laminar stress tensor is $\tau^{lam}_{ij} = 2\mu \widetilde{S}_{ij}^d$, with $\tilde{S}_{ij}^d$ the deviatoric part of the resolved strain rate tensor. The unresolved sub-grid scale (SGS) stress tensor is modeled using the Boussinesq assumption~\cite{boussinesq1877essai} as $\tau^{sgs}_{ij} = 2\mu_t \widetilde{S}_{ij}^d$, with $\mu_t$ the sub-grid turbulent viscosity, for which a model has to be provided. The so-called \textbf{$\sigma$}--model~\cite{nicoud2011using} is used, where $\mu_t$ reads:
\begin{equation}
    \mu_t=\rho (C_{\sigma}\Delta)^2 \mathcal{D}_{\sigma}(u)
\end{equation}
where $C_\sigma =1.35$ is the SGS model constant, $\Delta$ is the subgrid characteristic length scale and $\mathcal{D}_{\sigma}$ is a differential operator associated with the model, which reads:
\begin{equation}
    \mathcal{D}_\sigma(u)=\frac{\sigma_3(\sigma_1 - \sigma_2)(\sigma_2 - \sigma_3)}{\sigma_1^2}
\end{equation}
with $\sigma_1\geq\sigma_2\geq\sigma_3\geq0$ representing the three singular values of the local velocity gradient tensor. The \textbf{$\sigma$}--model has been built to yield zero SGS viscosity in situations where the flow is laminar, so that it is notably well adapted to situations where transition to turbulence occur at moderate Reynolds numbers~\cite{nicoud2018large,zmijanovic2017numerical}. This model was used in previous simulations from our group dedicated to jets and puffs flows in the context of disease transmission \cite{abkarian2020speech,Yang2020towards,mendez2023microscopic}. 

Massless Lagrangian particles are used to visualise and characterise the air flows released during nasal breathing. As these particles do not have inertia, they are simply displaced by the local flow velocity. The position of each tracer $\vec{X}_i$ is advected using the equation 

\begin{equation}
\frac{\mathrm{d}\mathbf{X}_i}{\mathrm{d}t} = \mathbf{U}_i(\mathbf{X}_i,t),
\label{eq:tracer}
\end{equation}
where $\mathbf{U}_i$ is the local instantaneous fluid velocity interpolated at the location of tracer $i$. 


\subsubsection*{Liquid phase}

As our aim is to study the transport of particles relevant to airborne transmission, we simulate droplets emitted at a diameter of a few tens of microns and less. Larger droplets would rapidly leave the exhaled cloud and fall to the ground, so that simulating them would not bring any interesting perspective to the present study. 

Expiratory droplets are modeled as spherical Lagrangian particles, tracked individually with a point-particle Lagrangian approach. This is a usual method in numerical studies relative to respiratory disease transmission \cite{balachandar2020host,chong2021extended,Wang2021short-range,Cortellessa2021close}. 
Their dynamics is governed by Newtonian mechanics and the size of the droplets can evolve due to evaporation. They are considered as sufficiently dispersed to neglect the particle-particle interactions. In addition, the low concentration of particles justifies a one-way approach, which would not be true to study coughs or sneezes.

The position of each particle is advanced by the kinetic equation
\begin{equation}
\frac{\mathrm{d}\mathbf{x}_p}{\mathrm{d}t} = \mathbf{u}_p,
\label{eq:transport}
\end{equation}
where $\mathbf{x}_p$ is the position of the particle and $\mathbf{u}_p$ its velocity.

The conservation of momentum is given by Newton's second law:

\begin{equation}
\frac{\mathrm{d}}{\mathrm{d}t}\left(m_p \mathbf{u}_p\right) = \mathbf{F}_p^G + \mathbf{F}_p^A,
\end{equation}

where $m_p$ is the mass of the particle, $\mathbf{F}_p^G$ is the buoyancy force and  $\mathbf{F}_p^D$ the drag force. The buoyancy force and drag force read:

\begin{equation}
\mathbf{F}_p^G = \left( \rho_p - \rho \right)\frac{\pi}{6}d_p^3\mathbf{g} \quad \quad {\rm and} \quad \quad \mathbf{F}_p^D = m_p\frac{1}{\tau_p}\left( \mathbf{u}_p - \mathbf{u} \right),
\end{equation}
where $\rho_p$ is the particle density, $\rho$ the local gas density, $d_p$ the particle diameter and $\mathbf{g}$ the gravitational acceleration.
$\tau_p$ is the characteristic drag time. It is modeled with the empirical correlation of Schiller and Naumann~\cite{naumann1935drag} for moderate values of the particle Reynolds number. This correlation approaches to the Stokes law at low Reynolds numbers. In this one-way coupling, particle drag is accounted for to compute the particle acceleration but is neglected from the fluid point of view. 

Droplets exchange heat with the gas phase, may also evaporate and the mass transfer model of Spalding is adopted~\cite{spalding1950combustion}. The Spalding model is based on the assumptions that the evaporating particles are isolated single-component spherical droplets of infinite thermal conductivity and negligible thermal diffusivity. The droplet surface is assumed to be at equilibrium with the surrounding gas, whose properties are assumed to be constant from the droplet surface to the far field.

The droplet mass transfer is resulting from the integration of the mass conservation law from the droplet to the far field~\cite{sirignano1983fuel}.
\begin{equation}\label{eq:droplet_mass_transfer}
\frac{\mathrm{d}m_p}{\mathrm{d}t}= - \pi d_p \rho D Sh \log(1 + B_M),
\end{equation}
where $\rm{B}_M$ is the Spalding mass number,  $D$ is the diffusivity and $Sh$ is the Sherwood number which represents the ratio of the convective mass transfer to the rate of diffusive mass transport.

The diameter evolution is obtained from the mass transfer equation (Eq.~[\ref{eq:droplet_mass_transfer}]). Its temporal evolution is a function of an evaporation characteristic time $\tau_m$ and the droplet initial diameter:
\begin{equation}\label{eq:temporal_dial_evol}
\frac{\mathrm{d}d^2_p}{\mathrm{d}t} = - \frac{d^2_{p,0}}{2 d_p \tau_m} \quad \text{ with } \quad \tau_m =\frac{\rho_pd_{p,0}^2Sc}{4Sh\mu_{1/3}\log(1+B_M)}. 
\end{equation}

The droplet temperature $T_p$ evolution is obtained by integrating the energy conservation from the droplet's surface to the far field, leading to:
\begin{equation}\label{eq:droplet_temperatureprogression}
\frac{\mathrm{d}T_p}{\mathrm{d}t} = - \frac{1}{\tau_h} \left (T_p - \left ( T_\infty - \frac{L_v B_T}{C_{p,1/3}} \right ) \right ),
\end{equation}
where $\tau_h$ is a thermal characteristic time, $T_{\infty}$ is the temperature of the surrounding gas, $B_T$ the Spalding thermal number, $L_v$ the latent heat and $C_{p,1/3}$ the heat capacity at a preponderated temperature $T_{1/3}$ used to characterize the intermediate state of the gas around the droplet. 

As droplets contain a solid fraction, they cannot evaporate completely. Let $V_{\text{init}}$ be the initial volume of the droplet when it is emitted. $V_{\text{non-volatile}}$ is the non-volatile volume contained in the droplet. We can estimate that the final diameter of the droplet $d_{\text{final}}$ after complete evaporation is related to the initial diameter $d_{\text{init}}$ by the relationship

\begin{equation}
\label{eq:f_final}
d_{\text{final}} \geq (V_{\text{non-volatile}}/V_{\text{init}})^{1/3} d_{\text{init}}.
\end{equation}

In reality, there is a relationship of inequality because the non-volatile fraction may retain some water: in particular, the final diameter is slightly dependent on the ambient relative humidity \cite{stadnytskyi2020airborne,marr2019mechanistic}. 
The parameter $\eta=d_{\text{final}}/d_{\text{initial}}$ controls the final size of the droplets after evaporation and therefore depends on the non-volatile fraction contained in each droplet, which is an unknown parameter and even varies from one time of day to another for each individual. In the proposed model, $\eta=1/3$ is considered as constant parameter for all droplets, meaning the final droplet size is approximately 33.3\% of its initial diameter after evaporation \cite{Netz2020physics}. This yields remaining volume proportion of non-volatile substances of 3.7\%.

For the smallest particles, of a micron and less in diameter, inertia is so small that 
they are essentially transported by the local air flow velocity \cite{Netz2020physics}. In addition, evaporation has a negligible effect on their dynamics, as they are already small. 
Their dynamics is thus well modeled by that of massless tracers during the few tens of seconds studied by simulation.   During nose breathing, emitted  droplets have a diameter of the order of a micron \cite{bagheri2023size}. As a consequence, in nose breathing flow simulations, only the transport of massless tracers is studied.
For the laugh flow simulations, particles are all treated with the full framework described by Eq.~[\ref{eq:transport}]-[\ref{eq:f_final}]. We limit the size of the injected droplets to the range [1;60] microns: this is justified by the fact that small particles have a small characteristic drag time $\tau_p$ that needs to be solved by our explicit method. Including droplets of 0.1 $\mu$m in diameter in the simulations would increase their cost with no practical interest, as they would  behave essentially the same as droplets of 1 $\mu$m in diameter over the time considered. In addition, as we are interested in the differences between the behavior of particles for airborne transmission, large droplets are not simulated. As shown by Fig. 3 of the manuscript, particles of initial diameter 60 $\mu$m are already almost absent of our analysis as they rapidly fall to the ground or deposit on the table. Particles with initial diameter in the range [60;100] microns should however be included in cases yielding faster evaporation or smaller diameters: dryer ambient air, smaller non-volatile fraction...

\subsection*{Computational domains}

The computational domain 
is a 3D parallelepiped of dimensions   \SI{4}{\meter} (width) x \SI{4}{\meter} (height) x  \SI{5}{\meter} (length) in the directions $x$, $y$ and $z$, respectively. A human manikin with mouth and nose openings is included in the domain. The face of the manikin is symmetric along the $x$-axis and directed toward the positive values of horizontal $z$-axis. It is positioned in a way that the jets or puffs may freely develop without interacting with the limits of the domain. $y$ is the vertical direction, pointing upward.

In the breathing flow simulations, twin breathing jets are released from two nostrils, symmetric along $x=0$. The cross-sectional area of a nostril is \SI{1.7}{\square\centi\metre}. The inflow is issued at the nostril exits.  A horizontal table 
is placed in front of the manikin with different vertical mouth-table separations of 20, 30 and 40 cm (nose-table separation distances of 22, 32 and 42 cm). 

For laughing flow, the inflow is released from the elliptic mouth exit of semi-axes $a_x=0.023$ m and $a_y=0.005$ m.  A horizontal table 
is placed in front of the manikin with a vertical distance of 40 cm from the mouth.

In both cases, the table is large enough so that the flow does not interact with the borders of the table.

\subsection*{Boundary conditions}

\begin{figure}
\centering
\includegraphics[width=\textwidth]{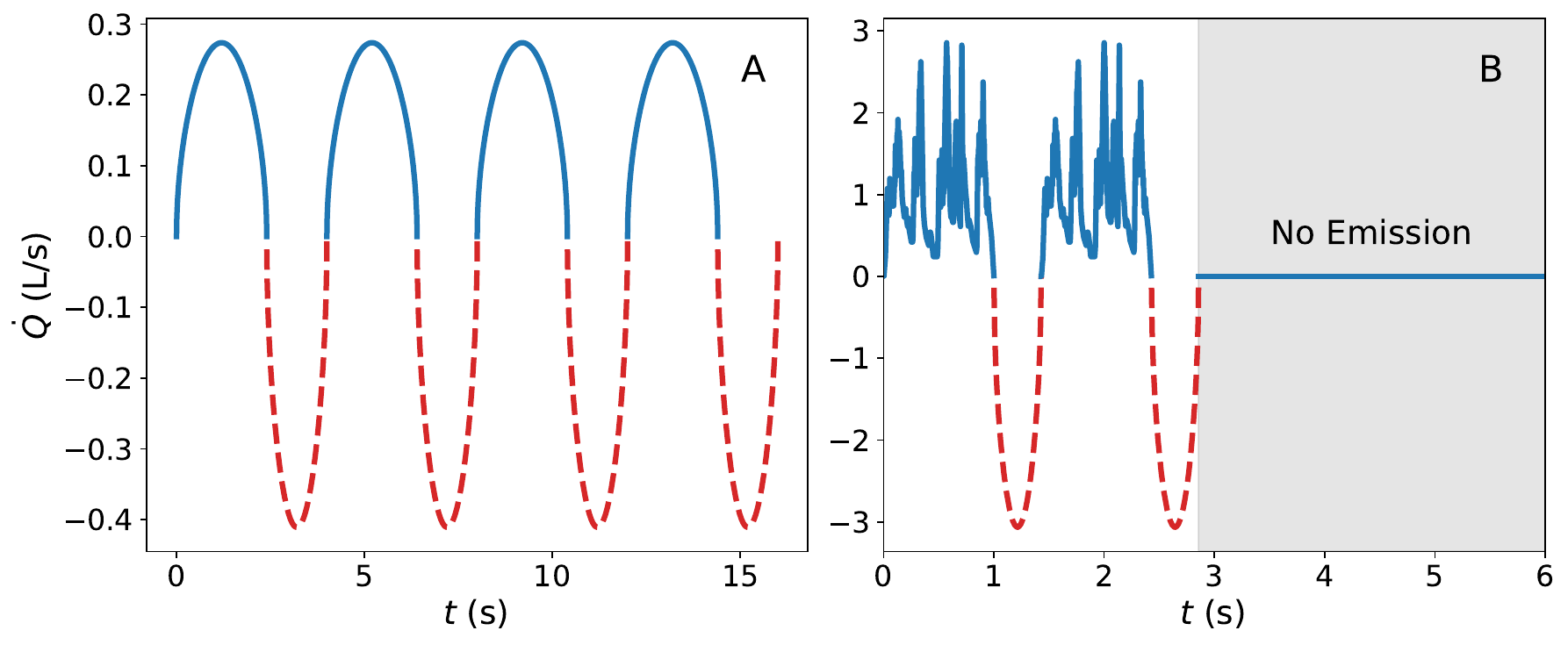}
\caption{Flow rate signals used in the simulations (solid lines: exhalation; dashed lines: inhalation): A) First four cycles of the signal for the nasal breathing flow with an exhaled volume of 1 L/cycle. The total duration of the simulations is 80 s, which corresponds to 20 cycles of 4~s, identical to the first four cycles displayed. Half of the flow rate is imposed at each nostril. B) First 6 seconds of the flow rate imposed for the laughing flow simulations. The signal in B is injected over 2 cycles of 1.5~s each, then the emission is set to zero for the rest of the simulation (between $t=3$~s and $t=125$~s). The exhaled volume is 1 L/cycle.}
\label{fig_SI:flowrates}

\end{figure}

The inflow for laughing and breathing flows is imposed as a uniform Dirichlet boundary condition at either the mouth or nostril exit. Flow rate signals are shown in Fig.~\ref{fig_SI:flowrates}. 

For breathing flows, the exhaled volume is 1 L per cycle/breath in the configurations of no-table, and for nose-table distances of 22 cm, 32 cm and 42 cm. The inflow signal with a duration of 4 seconds is divided into two phases of exhalation and inhalation. The duration of the phases is not equal (2.4 s for exhalation, 1.6 s for inhalation), but the exhaled and inhaled volumes are the same. The signals are multiplied by factor of 0.66 and 2.0 to modulate the flow rate to obtain lower (0.66 L/cycle) or higher (2 L/cycle) exhaled and inhaled volumes. The cycle is repeated continuously until the end of the  simulations. 
One jet is issued from each nostril, with half of the flow rate injected. The jets are inclined by an angle of $\SI{45}{\degree}$ downwards relative to $z$ in the ($z$,$y$) plane ; they are oriented at an angle of $\SI{25}{\degree}$ degrees with respect to $-y$, in the ($x$,$y$) plane (see Fig. 1 in the manuscript).
Massless Lagrangian tracers are also injected along with the exhaled flow. Their injection rate of is linearly proportional to the air flow-rate, fixed as 35,000 tracers per cycle or 1 L of exhaled air. The evolution of their position in time is resolved using Eq.~[\ref{eq:tracer}].

In simulations of laughing flow, the inflow is a time-varying, spatially uniform flow rate signal that contains jitter in amplitude, released from the mouth-exit. The duration of the exhalaation is 1 second with an exhaled volume of 1 L/cycle, followed by an inhalation of 0.5 s (and 1 L in inhaled volume). The signal is repeated 2 cycles, then the flow rate is set to zero until the end of the simulation. 
The inflow is seeded with spherical water droplets varying in diameter from 1 to \SI{60}{\micro\meter}, which are randomly injected at the mouth-exit. The volume proportion of ejected droplets to the gas phase is targeted to be constant at all times during the emission. Another imposed target is that the number of particles at the end of the emission follows a uniform distribution across the diameter range. This facilitates the statistical analysis of droplets that are widely dispersed throughout the ambient environment. Indeed, the aim here is not to inject a droplet distribution mimicking measurements (as in ref. \cite{chong2021extended} for instance), but to seed the flow with Lagrangian particles whose dynamics is representative of what may happen to particles in such a flow, and inject a large number to be able to readily build statistics. This is permitted by the assumption of one-way coupling. 16,500 particles are injected per cycle, so the total number of injected particles is around 33,000 at the end of the two cycles of injection mimicking the laugh.

The no-slip (zero velocity) Dirichlet boundary condition for velocity is imposed on the table and the head. The solid surfaces are modelled as adiabatic and impermeable boundaries, thus there is no mass or heat transfer from/to the air flow and ambient environment. In particular, this applies to the body of the emitter, so that thermal plume effects are neglected, as classically performed in current CFD studies for transport of pathogens. Given the conditions explored, the velocities in the region of interest are of the order of a few centimeters per second, which is much less than the order of magnitude of the velocities in the jets (a few meters per second). Note also experiments on manikins have shown the velocities in thermal convective boundary layer are decreased when the manikin is seated at a table \cite{Licina2014experimental}, the table screening the flow associated with the lower part of the body. 

\subsection*{Numerics}

The equations modelling the air motion and associated particle transport are solved using the YALES2BIO solver \cite{abkarian2020speech,Yang2020towards,mendez2023microscopic,zmijanovic2017numerical} which is derived from the YALES2 solver \cite{yales2023,moureau2011design,boulet2018modeling,grenouilloux2023toward}. The solver uses the projection method, initially introduced by Chorin \cite{chorin1968numerical} and subsequently modified by Kim and Moin \cite{kim1985application}, relying on the Helmholtz-Hodge decomposition. The prediction step is advanced in time by employing a TFV4A scheme, combining a 4th-order Runge-Kutta scheme with a 4th-order Lax-Wendroff-like scheme. The Deﬂated Preconditioned Conjugate Gradient algorithm (DPCG) is used for the correction step that involves solving a Poisson equation to determine pressure.  YALES2BIO was already used in the airborne transmission context \cite{abkarian2020speech,Yang2020towards,mendez2023microscopic}.

The convective time-step is fixed using the Courant–Friedrichs–Lewy (CFL) condition which is set to 0.4. Thermal diffusion is another constraint on the time-step, the associated viscous time-step is fixed according to Fourier number, which is set to 0.15.

To advance the position of tracers using Eq.~[\ref{eq:tracer}], a third-order Runge-Kutta scheme is employed. The velocity of tracers is determined through the interpolation of the local flow velocity at the tracer position.

\subsection*{Meshing}
A first tetrahedral mesh is generated in the 3D parallelepiped computational domain. Then, conformal meshing of the surface is obtained inside the in-house solver, which uses a grid cutting algorithm and a subsequent remeshing with the MMG3D mesh adaptation library \cite{dapogny2014three} to generate the starting grid \cite{grenouilloux2023toward}. That grid  has a grid size of 1 mm close to the face of the manikin and 5 mm and the remaining surfaces and is coarse far from the manikin. Due to the coarse grid in the far-field, the initial grid is not adapted to large-eddy simulations. It is refined on the fly, during the calculation, in regions where the vorticity norm is higher than a small threshold and where the air composition differs from that of the environment. The MMG3D library is used to adapt the grid \cite{grenouilloux2023toward}. In addition, the mesh is never coarsened during the simulation and a maximum gradient of mesh size is imposed to prevent grid discontinuities. The maximum target grid size is set to 8 mm. This yields final grids of 70 to 100 million elements depending on the case.

\begin{figure}
\centering
\includegraphics[width=\textwidth]{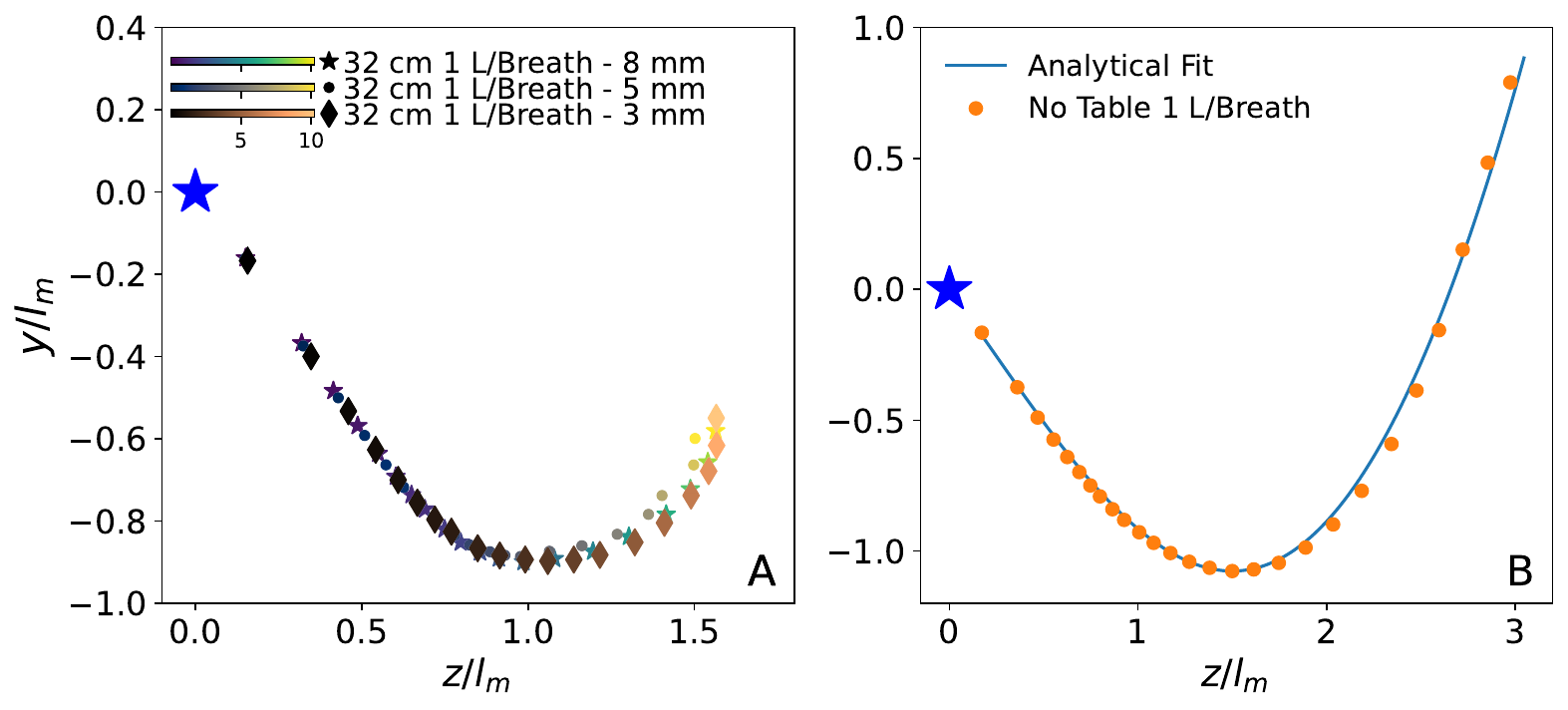}
\caption{A): Average trajectories of breathing jets with different target grid sizes in the bulk of the domain (3, 5 and \SI{8}{\mm}, color-coded by residence times of the massless tracers normalised by $l_m=D Fr$. Case with table place 32 cm below the nose. B) Non-dimensionalized trajectory of free buoyant-jet (`No Table'), compared with analytical model in the form of quartic polynomial proposed by \cite{papakonstantis2020simplified}. The polynomial coefficients are calculated by solving the set of equations (Eq. [9]-[14] in the cited paper), using the parameters $z_{min}$, $y_{min}$ obtained from the LES data.}
\label{fig_SI:trajectory_notable_model}

\end{figure}

In order to verify mesh convergence, simulations of the nose breathing flow are performed for the first 20 seconds in the case with a table-to-nose distance of 32 cm and an exhaled flow rate of 1 L/cycle. In Fig.~\ref{fig_SI:trajectory_notable_model}A, trajectories built from the Lagrangian tracers (which is extensively discussed in the manuscript) show that the flow is marginally impacted by the grid size. The grid used in the manuscript is thus shown to be sufficiently fine to explore effects on the trajectories.
Differences in the ascending portion are attributed to the limited integration time: here, trajectories being built from particles with a residence time between 0 and 12 s, discarding the solution of the first 8 s. In any case, the differences seen in Fig.~\ref{fig_SI:trajectory_notable_model}A due to the grid are smaller than those discussed in the article. 


\subsection*{Simulation cases}\label{SI:cases}

Table~\ref{table_SI:ref_table} reports the simulation parameters, inflow conditions and the trajectory characteristics of the breathing jets presented in the paper. The compositions of exhaled air and ambient air are provided in terms of mass fraction in Table~\ref{table_SI:ref_table_comp}. Relative humidity has been set to 90\% for the exhaled air. A temperature of 33$^\circ$ to 35$^\circ$ is generally used in models at the mouth exit \cite{chong2021extended,wang2022modelling,Cortellessa2021close}. Here, $T_{exh}=34^\circ$ is used at the mouth exit for the laughing flow. At exhalation, Ferron \textit{et al.} report slightly lower temperature in the nose than in the mouth \cite{ferron1988inhalation}, which led us to impose $T_{exh}=32^\circ$ at the nostrils for nose breathing. The ambient air is at $T_{amb}=25^\circ$ and 50\% relative humidity. 

\section*{Post-processing and complementary results}

\subsection*{Breathing jet trajectories}

The average trajectories of breathing jets $\bm{\vec{X}}(t_r)$ are calculated by averaging the positions of the $i$ tracers $\vec{x}$ exhaled from the two nostrils ($j=1,2$), using Eq.~[\ref{eq:trajectories}]:

\begin{equation}
    \bm{\vec{X}}(t_r)=\frac{1}{2N}\sum_{j=1}^{2} \sum_{i}^{N} \vec{x}_{i,j}(t_r),
    \label{eq:trajectories}
\end{equation}

\noindent $t_r$ is the residence time of tracers, which denotes the time since their injection. $2N$ is the total number of tracers considered in the computation of the trajectory. Summing of the two nostrils yields an average trajectory in the ($z$,$y$) plane. 
To obtain well-converged trajectories and discard the initial transient phase of the starting jets and wall-jets, we use tracers with injection time from 20 to 80 seconds. In addition, results for $t_r$ higher than 25 s are not displayed due to insufficient statistics.  Movie S4 shows the evolution of the exhaled particle cloud in the ambient, superimposed with the average trajectories of the jets. 

Figure~\ref{fig_SI:trajectory_notable_model}B reports the breathing jet average trajectory for the case without a table and a flow rate of 1 L/cycle and compares it with a quartic polynomial fit from the literature~\cite{papakonstantis2020simplified}. It shows that the expected trajectory of the free jet is well predicted with our large-eddy simulations, which confirms the quality of the simulations presented.

\subsection*{Laughing flow and particle deposition on the table}

Movie S5 shows an animation of various quantities of interest during the first 30 s of the laughing flow in the presence of a table located 40 cm below the mouth of the emitter: velocity, temperature, CO2 concentration and particles. Fictive heads are superimposed at the location where we have placed spherical probes to estimate the particles possibly inhaled by interlocutors. Those heads are not included in the CFD itself. Movie S5 clearly show a flow structure similar to Movie S3, with the impact on the table and the formation of a large vortical structure travelling along the surface. The bottom left panel shows that large particles essentially deposit on the table at the impact and are essentially absent from the rising cloud.

\begin{figure}
\centering
\includegraphics[width=\textwidth]{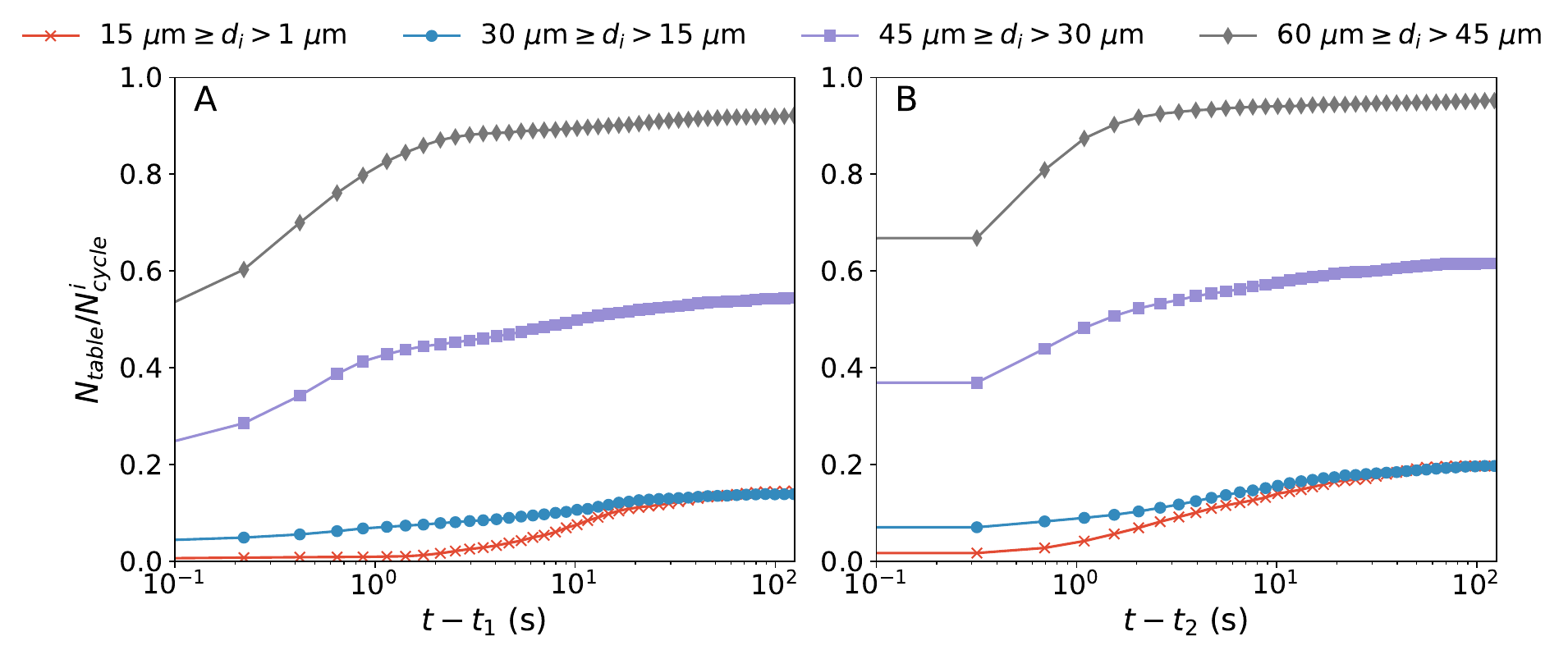}
\caption{Time evolution of particle deposition on the table for the puffs released during laughter: (A) first cycle; (B) second cycle. Droplets of initial diameter ranging from \SI{1}{\micro\meter} to \SI{60}{\micro\meter} are divided into 4 groups, as indicated in the legend. Note that the size distribution of emitted particles is uniform. $t$ denotes the physical time with $t_i$ representing the time of the end of emission of cycle $i$.}
\label{fig_SI:droplet_deposition_time}
\end{figure}

Figure~\ref{fig_SI:droplet_deposition_time} shows the particle deposition on the table in the laughing flow, plotted versus time. The particles are grouped based on the injection cycle during which they are released and the initial diameter of the particles (droplets) at the emission. As shown in Fig. 4B of the manuscript, almost all of the large particles (45-\SI{60}{\micro\meter}) are deposited on the table soon after the end of the injection, due to inertial impact. Figures~\ref{fig_SI:droplet_deposition_time}A,B show that the deposition occurs rapidly, with 90\% of the particles in this diameter range being deposited on the surface within 1 second after the end of the emission. On the other hand, gravitational sedimentation progressively becomes important for particles with smaller diameter. As the sedimentation velocity of these particles is proportional to their diameter, they can remain suspended in the released warm pocket of air for longer periods of time, resulting in less deposition on the surface. 

\begin{figure}
\centering
\includegraphics[width=\textwidth]{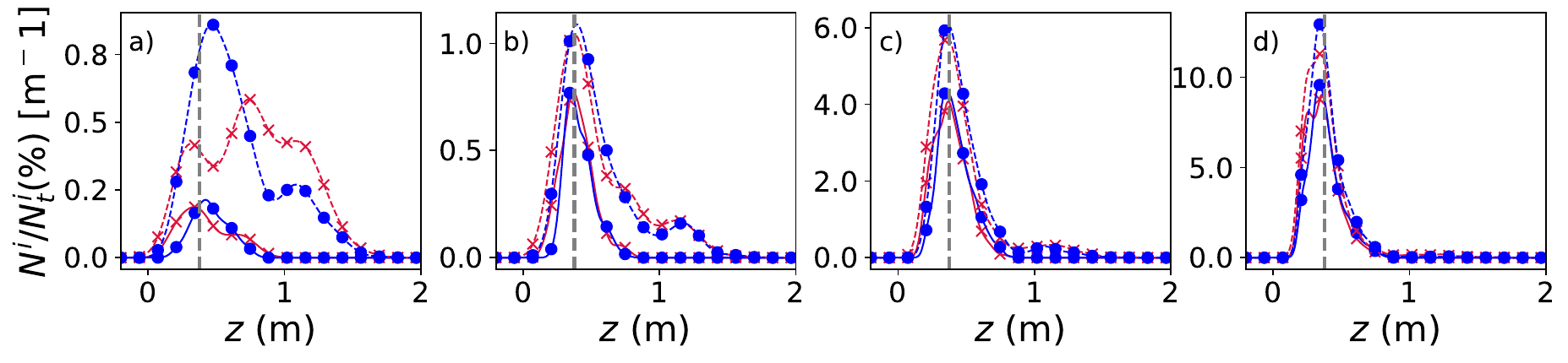}
\caption{Particle deposition profiles on the table for the puffs released during laughter. Particles are grouped into 4 bins based on their initial diameter, i.e a) \SI{1}{}-\SI{15}{\micro\metre}, b) \SI{15}{}-\SI{30}{\micro\metre},  c) \SI{30}{}-\SI{45}{\micro\metre}, d) \SI{45}{\micro\metre}-\SI{60}{\micro\metre}. 
    Lines and markers are color-coded by the injection cycle: first cycle (red circles); second cycle (blue crosses). Solid and dashed lines, respectively, denote 5 and 125 s of simulated time. The vertical dashed line is the location of the geometric impinging point of the puffs.}
\label{fig_SI:droplet_deposition_profile}
\end{figure}

The prevalence of these deposition mechanisms is also supported by the deposition profiles shown in Figure~\ref{fig_SI:droplet_deposition_profile}. The peak deposition densities for large particles (Fig.~\ref{fig_SI:droplet_deposition_profile}c,d) injected during both cycles are located near the geometrical impinging point of the exhaled air. As the particle diameter decreases, the contribution of gravity becomes visible, which is corroborated by the second peak in the deposition profile (Fig.~\ref{fig_SI:droplet_deposition_profile}a,b). As the turbulent puff moves downstream, it sheds small particles which eventually sediment on the surface. The observable contrast between the first and second cycle in Fig.~\ref{fig_SI:droplet_deposition_profile}a can be attributed to the fact that second puff released almost immediately after the first puff feeds momentum to the latter which carry smallest particles further downstream. For larger particles, differences between the two cycles are minimal.  

\section*{Movies: visualizations of the breathing and laughing flows}

\subsection*{Illustrative experiments}

Experimental illustrations of the exhaled flows are presented. They should not be considered as validation cases, as the operating conditions are not controlled in those experiments: the room and exhaled flow temperatures, flow rates, volume, upper airways shapes and flow angles have not be measured, so that quantitative comparison is not possible. However, these experiments illustrate the type of flow motions studied in the simulations and display a number of key features discussed in the article. All movies presented in this document can be accessed \href{https://www.dropbox.com/scl/fo/34w7kwrntrwpbn0ojptkv/AOEw0yjkOcmhlramN0od8As?rlkey=056wh58r9xzqfhgb0vp2rnjkb&e=1&st=qab3xjmg&dl=0}{\textcolor{blue}{at this link}}.

Nose breathing flow is illustrated thanks to Movies S1 and S2, showing laser-sheet flow visualisations of a subject performing a deep exhalation through the nose. Only one cycle is visible. In Movie S1 the subject is oriented so that the exhaled flow is in the laser sheet plane. The impact on the table, the clinging, then the vertical movement of the plume are observable. In movie S2, the subject is located behind the laser sheet and faces it. It illustrates the lateral motion associated with the angle of the exhaled nostril jets.

In both movies, the environment is seeded with droplets with visible flow structures to ease the observation of motions at the naked eye. The exhaled flow is characterized by a lower concentration of droplets (darker region), faster motions and smaller length scales than in the ambient air. Due to the position of the subject with respect to the laser, the flow at the nose is not visible.

\begin{figure}
\centering
\includegraphics[width=\textwidth]{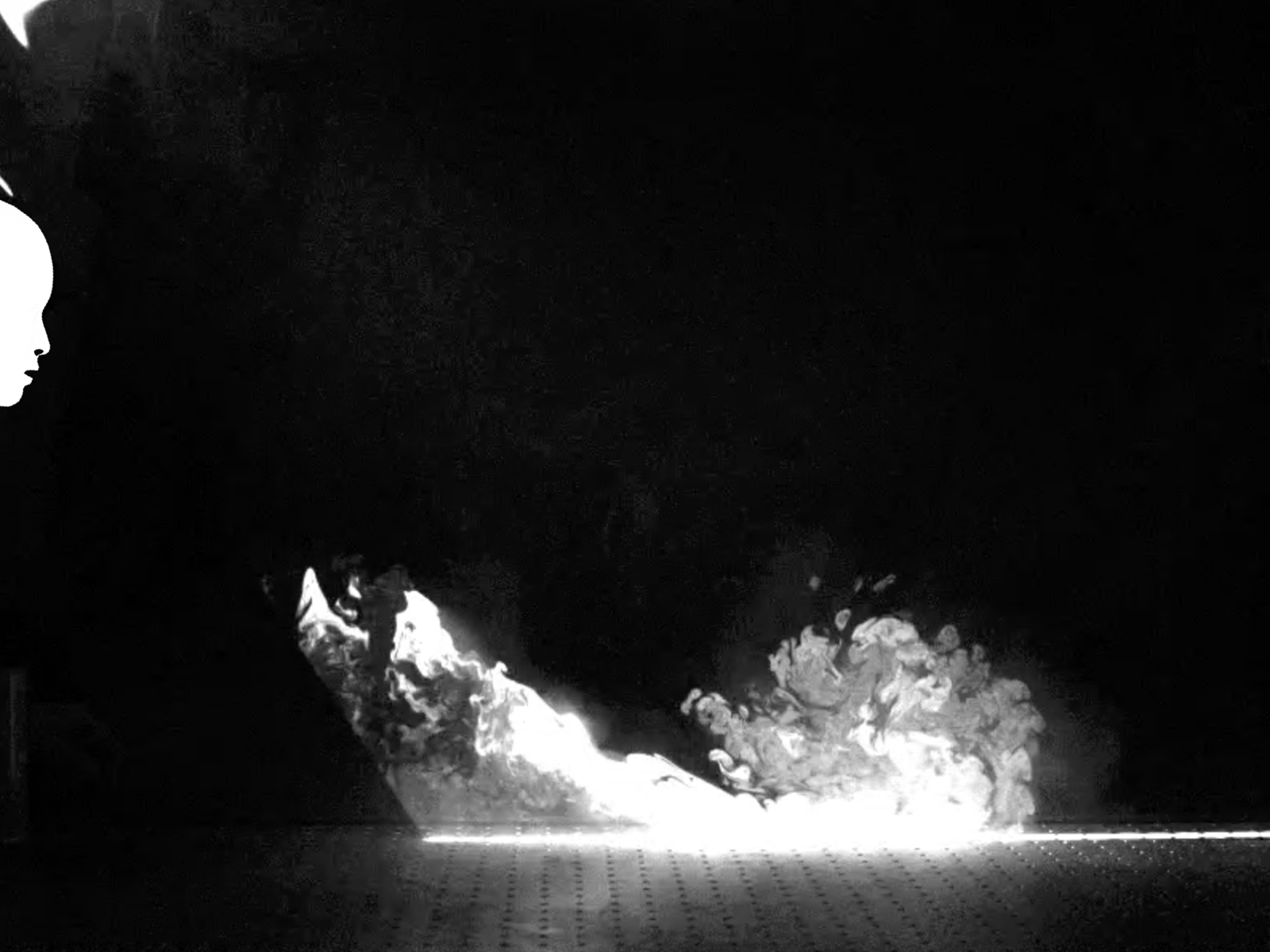}
\caption{Snapshot of a subject laughing in front of a table. The exhaled flow drags a fog of droplets in the laser sheet, which enables flow visualization. Due to the position of the subject with respect to the laser, the flow at the mouth exit is not visible.}
\label{fig_SI:laugh_manouk}
\end{figure}

Movie S3 presents a series of laughs in front of a table. Fig.~\ref{fig_SI:laugh_manouk} displays one snapshot of the movie on which the location of the laughing person, which is barely visible in the movie, is here clearly displayed. Contrary to the previous movies for the breathing flow, the droplets seed the environment near the subject's mouth, so that they are dragged with the laughing flow and enable to visualize the exhaled flow \cite{abkarian2020speech}. 

\subsection*{Numerical simulations}

Two movies are presented to visualize the effect of the table on the development of the breathing flow (movie S4) and laughing flow (movie S5).

\vfill

\pagebreak

\pagebreak

\pagebreak

\pagebreak

\pagebreak
\clearpage

\begin{table}
\scriptsize
\centering

\newcolumntype{L}{S[table-format=1.2,round-mode=places, round-precision=2]}
\newcolumntype{M}{S[table-format=1.0,round-mode=places, round-precision=0]}
\begin{tabular}{cccLLcSSLLL} 

    & {Case} & {$Q$ (L/cycle)} & {$U_{0,max}$ (m/s)} & {$\overbar{U}_0$ (m/s)} & {$\overbar{Re}(Re_{max})$} & {$\overbar{Fr}(Fr_{max})$} & {$h_{t}$ (cm)} & {$h_{\text{t}}/l_m$} & {$y_{{min}}/l_m$} & {$z_{@y_{min}}/l_m$} \\ \midrule
    {Breathing} & {No Table} & 1.0 & 2.337 & 1.780 & 1449  & 24.87 & {-} & {-} & 1.076 & 1.360 \\
    & {20 cm}    & 1.0 & 2.337 & 1.780 & 1449 & 24.87& 22.2 & 0.730 & 0.605 & 0.714 \\
    & {30 cm}    & 1.0 & 2.337 & 1.780 & 1449  & 24.87& 32.2 & 1.058 & 0.887 & 1.095\\
    & {40 cm}    & 1.0 & 2.337 & 1.780 & 1449  &  24.87& 42.2 & 1.387  & 1.119 & 1.410  \\ 
    & {30 cm - Qx0.666}  & 0.66   & 1.557 & 1.187   & 965  & 16.56 & 32.2 & 1.591 & 1.146 &  1.529  \\
    & {30 cm - Qx2}  & 2.0  & 4.675 & 3.563  & 2899 &  49.74 & 32.2 & 0.530 & 0.439 & 0.527 \\ \midrule
    Laughing & {No Table}  & 1.0  & 7.061 & 2.34  & 3900 (11750) & 23 (69) & {-}  & {-} & {-} & {-}   \\
    & {Table}  & 1.0  & 7.061 & 2.34  & 3900 (11750) & 23 (69) & {40}  & {-} & {-} & {-}    \\ 
\end{tabular}

\caption{Summary of characteristics of numerical simulations presented in the paper. The inflow signals are presented in Fig.~\ref{fig_SI:flowrates}. $U_{0,max}$ and $\overbar{U}_0$: Maximum and mean velocity of the inflow over one cycle of exhalation; $\overbar{Re}$ and $\overbar{Fr}$: The mean Reynolds and Froude numbers over the exhalation time of one cycle, for the laughing flow simulations ${Re_{max}}$ and ${Fr_{max}}$ denote the maximum Reynolds and Froude numbers reached during the emission; $h_t$: vertical distance of the table to the emitter; $l_m$: characteristic length of the buoyant jet; $y_{min}/l_m$ and $z_{@y_{min}}/l_m$: maximum vertical penetration of the breathing jets and the associated horizontal location, respectively.}

\label{table_SI:ref_table}
\end{table}

\begin{table}
\scriptsize
\centering

\newcolumntype{L}{S[table-format=1.4,round-mode=places, round-precision=4]}
\newcolumntype{M}{S[table-format=1.0,round-mode=places, round-precision=0]}

\begin{tabular}{ccMLLLLL} 

{Configuration} & {Temperature} & {RH (\%)} & {$Y_{O_2}$} & {$Y_{N_2}$} & {$Y_{Ar}$} & {$Y_{H_2O}$} & {$Y_{CO_2}$}  \\ \midrule

\multirow{ 2}{*}{Breathing} & {$T_{exh}$= \SI{32}{\celsius}} & 90 & 0.1678 & 0.7196 & 0.0123 &	0. 0274 &	0.0729 \\
& {$T_{amb}$= \SI{25}{\celsius}} & 50 & 0.2291	& 0.7478& 0.0127& 0.0099& 0.0005 \\  \midrule

\multirow{ 2}{*}{Laughing} & {$T_{exh}$= \SI{34}{\celsius}} & 90 & 0.16721 & 0.71703 & 0.01222 & 0.0309	& 0.07264 \\
& {$T_{amb}$= \SI{25}{\celsius}} & 50 & 0.2291	& 0.7478& 0.0127& 0.0099& 0.0005 \\  

\end{tabular}
\caption{Characteristics of the exhaled and ambient air: $T_{exh}$: Exhaled air temperature; $T_{amb}$: Ambient air temperature;  RH: Relative humidity. Mass fraction of  oxygen ({$Y_{O_2}$}), nitrogen ({$Y_{N_2}$}), argon ({$Y_{Ar}$}), water vapor ({$Y_{H_2O}$}) and carbon dioxide ({$Y_{CO_2}$}).}
\label{table_SI:ref_table_comp}
\end{table}

\pagebreak
\clearpage


\movie{Laser-sheet flow visualization of the flow exhaled during nasal breathing. The subject is oriented so that the exhaled flow is aligned with the laser sheet. The frame rate is resampled to play at real time.}

\movie{Laser-sheet flow visualisation of the flow exhaled during nasal breathing. The subject is facing the laser sheet, a few tens of centimeters away. The flow from the right nostril is visible. The frame rate is resampled to play at real time.}

\movie{Laser-sheet flow visualisation of puffs expelled during a bout of strong laughter and interaction with the table. A large vortical structure travelling along the table is visible, similar to the one observed in the simulations (Movie S5).}

\movie{Time evolution of particle clouds (tracers) for breathing jets, colored in blue. In clockwise order: A) No Table; B) 40 cm; C) 30 cm; D) 20 cm. The time-averaged trajectories of jets, colored by the residence time, are superimposed on the clouds. The cases displayed are the ones used in Fig. 2A, 2B and 2D of the manuscript.}

\movie{Simulated velocity,  scalar fields and particle transport following the expulsion of puffs during a typical laughter: Top-Left: Flow velocity; Top-right: CO$_2$ (PPM) Bottom-left: Temperature; Bottom-right: Instantaneous particle diameter after their emission as droplet at the mouth. Three heads facing the emitter are representative and are not included in the simulation. They mimick the 3 spheres used in the main paper to estimate particle inhalation in Fig. 4A of the manuscript. The case displayed is the same as the one used in Fig. 3A-F and Fig. 4A-B with Table in the main manuscript. Only the first 30 s of the simulation are shown.}

\clearpage
\bibliographystyle{unsrt}
\bibliography{covid_biblio}